\documentclass[aps,showpacs,prb,twocolumn,supperscriptaddress]{revtex4-1}
\usepackage{graphicx} 
\usepackage{dcolumn} 
\usepackage{bm} 
\usepackage{graphicx,color,epsfig,rotate}
\usepackage{amsmath,amssymb}
\usepackage{pifont}

\makeatletter
\newcommand{\Rmnum}[1]{\expandafter\@slowromancap\romannumeral #1@}
\makeatother
\newcommand{\dbar}{d\mkern-5mu\mathchar'26}

\begin{document}

\title{Re-equilibration after quenches in athermal martensites: \\Conversion-delays for vapour to liquid domain-wall phases}
\author{N. Shankaraiah$^{1,2}$, K.P.N. Murthy$^2$, T. Lookman$^3$ and S.R. Shenoy$^4$}
\affiliation{$^1$ School of Physical Sciences, Jawaharlal Nehru University, New Delhi 110067, India\\
$^2$School of Physics, University of Hyderabad, Hyderabad 500046, India \\ 
$^3$Theoretical Division, Los Alamos National Laboratory,  NM 87545, USA \\
$^4$ TIFR-Centre for Interdisciplinary Sciences, TIFR-Hyderabad, Hyderabad 500075, India}

\date{\today}

\begin{abstract}

Entropy barriers and ageing states appear in  martensitic structural-transition models,  slowly re-equilibrating after temperature quenches, under Monte Carlo dynamics. Concepts from protein folding and ageing harmonic oscillators turn out to be useful in understanding these nonequilibrium evolutions.
We show how the athermal, non-activated delay time for seeded parent-phase austenite to  convert to  product-phase martensite, arises from an identified  entropy barrier  in Fourier space.  In an ageing state of low Monte Carlo acceptances, the strain structure factor makes  constant-energy searches for rare  pathways, to enter a Brillouin zone  `golf hole' enclosing negative energy states, and to suddenly release entropically trapped stresses. In this context,  a stress-dependent effective temperature can be  defined,  that  re-equilibrates to the quenched bath  temperature.

\end{abstract}

\vskip 0.5truecm

\pacs{64.70.Q-, 81.30.Kf, 64.70.K-, 87.15.Cc}

\maketitle

\section{Introduction}

Ageing states of  glassy systems are a longstanding puzzle \cite{R1}. They are pictured as being metastably trapped in a multi-valley free-energy landscape in configuration space.   Their  re-equilibration times $\{ t \sim e^{\Delta F/  T}\}$ (or inverse rates)   to find the global minimum, depend on the set of 
free energy barriers $\{\Delta F = \Delta U - T \Delta S\}$ between basins. Although  both  energy barriers $\Delta U$ and entropy barriers $|\Delta S| = -\Delta S$ can contribute,  these two delay sources are distinct. Energy barriers  require thermal  activation,  with Arrhenius rates as Boltzmann factors ${t}^{-1} \sim e^{-\Delta U / T}$, while entropy barriers do not have such inverse temperatures in the exponent.  Hence if  deep quenches  face unsurmountable energy barriers, $t \sim e^{\Delta U/ T} >>1$, then  re-equilibrations could be  dominated by the entropy barrier delays $t \sim e^{|\Delta S|}$, coming from  searches  at almost constant energy $\Delta U/T <<1$, for rare canyon-like pathways connecting the basins. 

The folding of proteins involve such entropic searches. The folded state of the macromolecule  is one of a huge number of configurations, and a random search would take a very long time. Like  a  {\it golf hole}  in a flat surface, there are many ways of failing, and only a few of succeeding: this is the entropy barrier. The observed  rapid folding is understood through the concept of a guiding {\it funnel} leading to the folded state \cite{R2,R3,R4}. Protein folding toy models of a Brownian particle  searching outside a golf hole  (`unfolded state') for  a funnel inside it (`folded state'), also find entropy barriers, at the golf  hole edges \cite{R3}. 

Entropy barriers and ageing behaviour,  arise in several  non-interacting statistical models. Backgammon models of  freely hopping particles \cite{R5},  and ageing  harmonic oscillators under Monte Carlo dynamics \cite{R6}, both without energy barriers, can show  glassy  behaviour  from entropy barriers alone.  In an ageing state of harmonic oscillators, the  MC {\it  acceptance fractions} in a sweep over all sites are very small, and decrease  monotonically and slowly with time. Kinetically constrained models with trivial statics, but with  imposed constraint on the spin-flip dynamics, can exhibit slow relaxations, and glass-like  freezing at `entropy catastrophes' \cite{R7}. Glass-like quasi-steady states are found in models with infinite-range or power law interactions \cite{R8}. 

 We find that models of  structural transitions, used to study unusual quench responses of athermal  martensites, can also shed light on such  general issues of nonequilibrium statistical mechanics. 

Martensitic transitions can have unusual dynamics; have naturally non-uniform states and power law interactions; and can be a rich source of interesting nonequilibrium models \cite{R9,R10,R11,R12}.  These structural transitions have  strain-tensor components as the order parameters, with parent-phase high-symmetry `austenite' converting  to different variants of lower-symmetry product-phase `martensite', separated by elastic  domain walls. These  twin-boundaries are oriented  in preferred crystallographic directions by anisotropic  power law interactions, arising from generic St Venant Compatibility constraints, that ensure lattice integrity \cite{R11}. 

Martensitic materials are classified as  `isothermal', with activated, slow conversions; or `athermal', with very rapid conversions when quenched below a martensite start temperature, and with no conversions,  above it \cite{R9}. However in puzzling experiments, some athermal  materials convert even  {\it above} the rapid-conversion temperature, after  delays  of  thousands of seconds\cite{R10}.
 
Monte Carlo (MC) simulations  in athermal parameter regimes, of a square-to-rectangle (or a 2D version of the tetragonal-to-orthorhombic) transition, have considered re-equilibration after a  temperature quench \cite{R12}, with continuous strains  represented as in various contexts by discrete-strain pseudo-spins\cite{R13,R14}. For a square-cell austenite converting to one of  two possible  rectangle-cell variants of martensite, the pseudospin can take on three values of $0, \pm1$, as in a Blume-Capel model \cite{R15}.

For athermal-regime parameters \cite{R16}, dilute martensitic seeds in austenite of initial fraction $n_m (0)<<1$,  are  systematically quenched to well below the scaled Landau temperature $T=T_0=1$. They quickly form a martensitic droplet, that searches for an  autocatalytic twinning channel \cite{R17}, and finds it after an austenite-martensite  conversion delay $t_m$,  when there is a sharp rise of the martensite fraction towards unity. The conversion delays can be very long or very short, depending on temperature. The simulations\cite{R12} correctly model experiments \cite{R10}, showing very  fast   (slow) conversions at low (high) $T$. In this paper, we provide an understanding of these martensitic re-equilibrations, that  are also very relevant for glassy systems. 

The conversion times are  insensitive to energy-barrier scales, and therefore can only arise\cite{R12}  from {\it entropy barriers} $t_m \sim e^{|\Delta S|}$.  What  is the nature of the  entropy barrier  that blocks  immediate access to states of  manifestly lower energy? How can the barrier crossings at nearby temperatures,  be very fast, or very slow?

 We answer these questions, using concepts from protein folding and ageing harmonic oscillators. The natural description is in Fourier space. The initially isotropic {\it strain structure factor} makes a delay-inducing search on a constant-energy surface, for an anisotropic  `golf hole' in the Brillouin zone. The golf hole is bounded by the $\vec k$-space line where the energy spectrum of martensitic textures vanishes. We identify the distortion pathways for the structure factor to cross the entropy barrier and enter the negative-energy funnel region, when there is a sudden spike in the MC acceptances, with trapped stress released as heat. A strain-related effective temperature can be defined.
  
  In Section II we outline the pseudospin model used previously. In Section III we analyse the textural evolution in Fourier space, and in Section IV understand the crossing of the vapour-liquid entropy barrier.  In Section V we consider domain-wall thermodynamics, with details in the Appendix. Finally Section VI has a summary and conclusions. 
 
\section{The pseudospin model}

 The strain pseudospin \cite{R12,R14}  Hamiltonian is derived from a free energy  $F (e)$ in the order parameter strains $e(\vec r)$,  evaluated at the Landau minima through             $e \rightarrow {\bar \varepsilon} (T) S(\vec r)$. Here ${\bar \varepsilon} (T)$ is the strain magnitude at the minima; and the pseudospin locates the triple-well minima as                         $S (\vec r)=0, \pm 1$. Thus $F({\bar \varepsilon} S) \equiv H (S)= H_L + H_G + H_C$, where  the Landau term is  $H_L \sim \sum  g_L(T) S^2 (\vec r)$; the Ginzburg term is $H_G \sim \sum  {\xi_0}^2 ({ \vec \Delta} S )^2 $; and the St Venant  Compatibility  term is
 $H_C \sim (A_1/2)\sum  V ({\vec r} - {\vec r^{'}} ) S (\vec r) S  ({\vec r}^{'})$. The interaction is an anisotropic power law in the separation $R$  that is scale-free, $V(\lambda \vec R) \sim V(\vec R)/ \lambda^d$, with a spatial average that is zero. Here  ${\vec \Delta}$ is a difference operator on the square reference lattice. 
  
  In Fourier space ${\vec \Delta} \rightarrow i {\vec K}$ where  $K_\mu = 2 \sin( k_\mu /2)$. 
 The  Hamiltonian is diagonal in Fourier space, and is  {\it formally}  that of  ${\vec k}$-labelled  inhomogeneous oscillators \cite{R6,R18}.  
 The Hamiltonian energy of a pseudospin texture from MC dynamics at a time $ t$ is
 $$H (t)   = \frac{K_0}{2} \sum_{\vec k}  \epsilon (\vec k) |S (\vec k, t)|^2,~~(2.1a)$$
with   a dimensionless martensitic-strain spectrum 
$$\epsilon (\vec k) \equiv  g_L(T) +  \xi_0 ^2 { \vec K}^2   + \frac{A_1}{2} (1 - \delta_{{\vec k}, 0}){\tilde V} ({\vec k}),~(2.1b)$$
where the (positive) interaction  kernel  is \cite{R12}
$${\tilde V}(\vec k) =  (K_x ^2 - K_y ^2)^2 / [ K^4 + (8 A_1/ A_3) (K_x K_y)^2 ].~(2.1c)$$
 The kernel ${\tilde V} \sim \cos ^2 2\theta = (1+ \cos 4 \theta)/2$ vanishes along favoured $\theta = \pm \pi/4$ or $k_x = \pm  k_y$~~ diagonal directions, and so a nonzero  
contribution  ~$H_C \sim \sum {\tilde V }(\vec k) |S(\vec k,t)|^2 \neq 0$ is a domain-wall {\it mis}-orientation energy. The Fourier kernel average over the Brillouin zone (BZ) is ${\bar V} \equiv [{\tilde V}] \simeq 0.3$.  
 
Here \cite{R16}$ A_1, A_3$ are elastic constants; $E_0$ is an elastic energy per unit cell; and  $K_0 (T) \equiv 2 E_0 {\bar \varepsilon}^2 (T)$, where  ${\bar \varepsilon} (T) = [ \frac{2}{3} \{ 1 + \sqrt{ 1 -  3 \tau /4} \} ]^{1/2}$. The Hamiltonian energy for the $\vec k =0$ uniform state is  just the Landau energy  $H = N E_0 f_L \sum_{\vec r} S^2(\vec r)$, where  $f_L \equiv \bar{\varepsilon}^2 g_L$ and $g_L (T)  \equiv  \tau - 1  + ( {\bar \varepsilon}^2  - 1)^2 < 0$,   favouring martensite for $\tau < 1$. Here the scaled  temperature 
$$\tau (T)\equiv  ( T - T_c )/( T_0 - T_c) ~~~(2.2)$$
 at  the first-order Landau transition temperature $T_0$ is unity $\tau(T_0) =1$; while at the Landau spinodal $T_c$ where metastable \cite{R19} austenite becomes unstable $\tau (T_c) =0$. Notice  that $H$ depends on the quenched  {\it bath}  temperature, as a  bulk pseudo-spin can flip to equilibrate to $T$,   in a single $t \rightarrow t+1$ increment. The Hamiltonian $H(t)$  then describes the slower pattern evolutions of domain walls separating the $S= \pm 1$ variants; or separating one variant and the $S=0$ austenite. In the glass terminology,  the domain wall patterns are  `inherent structures'  at landscape minima \cite{R1,R4}. Similar models coupling strain to magnetisation or to random disorder  would be relevant for other glassy phenomena \cite{R9}.
 
Powerlaw {\it isotropic} interactions $\sim 1/R^\alpha$  with $\alpha < d$, with divergent spatial averages, show unusual statistical behaviour, with non-extensive thermodynamic functions;  inequivalence of ensembles; and ergodicity breaking of quasi-stationary states whose lifetimes can diverge with system size\cite{R8}. The $\alpha =d$ `marginal'  case has logarithmically divergent averages. In our case, the Compatibility interaction between the order parameters arises from minimizing non-order parameter strains subject to a constraint  connecting derivatives of all strains \cite{R11, R14}. Thus the $\vec k =0$ interaction contribution is zero as in (2.1c). The spatial average then vanishes, ~~~ $\sum_{\vec R} V(\vec R)/L^d \sim (1- \delta_{\vec k,0}  )\delta_{\vec k,0} =0$, and so this is a {\it sub}-marginal case. Nonetheless, we find  finite incubation delays from  finite entropy barriers.
\begin{figure}[h]
\begin{center}
 \includegraphics[height=6.2cm, width=8.3cm]{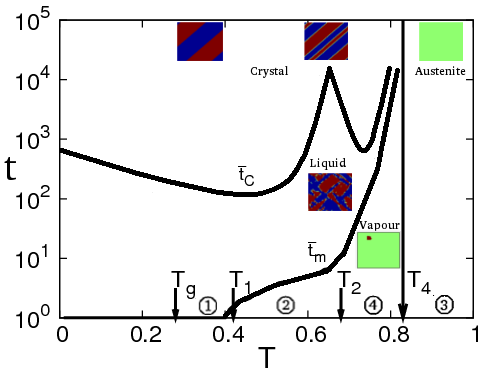}
\caption{{\it Temperature-Time-Transformation:} Log-linear plots of  mean martensitic conversion times ${\bar t}_m (T)$ and domain-wall orientation times  ${\bar t}_C (T)$ versus quench temperatures $T$. Pictures are snapshots of  domain-wall 'vapour' ($\bar t_{m} > t$), 'liquid' ($\bar t_{C} > t > \bar t_{m}$) and 'crystal' ($t > \bar t_{C})$. Quench regions are indicated: Region 4  with $T_4 > T > T_2$;  Region 2  with $T_2 > T > T_1$; and Region  1 with $ T_1 > T > T_g$. }
\label{Fig.1}
\end{center}
\end{figure}

Re-equilibration under a quench-and-hold protocol\cite{ R16}, is described by the dynamic structure factor
 
 $$\rho (\vec k, t) \equiv |S(\vec k, t)|^2 ,~~~(2.3)$$  
and its BZ average is the martensite fraction:  
 $$\sum_{\vec k} \rho (\vec k, t) /N = \sum_{\vec r} S^2 (\vec r, t) / N =  n_m (t).~~(2.4) $$

Fig 1 shows schematic curves from the data\cite{R12,R16}, of  the Temperature-Time-Transformation (TTT) evolution of the  seeded system, quenched to a temperature $T$.The free-running system is monitored as it passes through domain wall phases of vapour, liquid and crystal. The martensitic conversion time $t_m (t)$  where $n_m (t_m) =0.5$, is also the TTT phase boundary between vapour and liquid. There is also another time $t_C (T)$ for the orientation of domain walls, that will be considered elsewhere. 

We will consider  three  temperatures in three distinct temperature Regions \cite{R12}.  For $T_4 > T = 0.76 > T_2$  in Region 4, the re-equilibration is dominated by the vapour-to-liquid or conversion delay\cite{R20} $t_m (T)$.  For $T_2 > T= 0.55 > T_1$ in Region 2, the total delay also has contributions from the liquid-to-crystal delay $t_C (T)$. Finally, for $T_1 > T = 0.4$ in Region 1, the conversion time is negligible, and the domain-wall orientation time $t_C (T)$ dominates. In Region 3, even though the quenched temperature is still below the Landau transition temperature $ T_0=1 > T> T_4 =0.824$, the small, dilute  martensitic seeds disappear into surrounding metastable austenite, and are not re-nucleated. Hence in Region 4 we study a $T=0.76$ sufficiently below $T_4$, for a reasonable number of  the $N_{ runs} =100$ to convert, in a reasonable time.  

\section{Textural evolution in Fourier space}

Fig 2a)  shows that for a quench to $T=0.76$ the Hamiltonian energy of (2.1) is nearly flat, with $H(t) \simeq 0$, up to $t \sim t_{Sm}$. It starts to go negative at $t \sim t_{1} \sim 0.85~ t_m$; and falls rapidly at $t=t_m$. The fall slows down at $t  \sim  t_{SC}$,  and finally  flattens at   $t \sim t_C$. By contrast, for a quench to $T =0.4$ in Region 1, the  energy drops almost immediately.  Fig 2b) shows the austenite fraction $n_0 (t) \equiv 1 - n_m (t)$ behaves similarly, incubating at $n_0 (t) = n_0 (0) \simeq 1$ for $T=0.76$ before falling; while being expelled immediately, for $T =0.4$.
 
 \begin{figure}[h]
\begin{center}
\includegraphics[height=3.3cm, width=8.7cm]{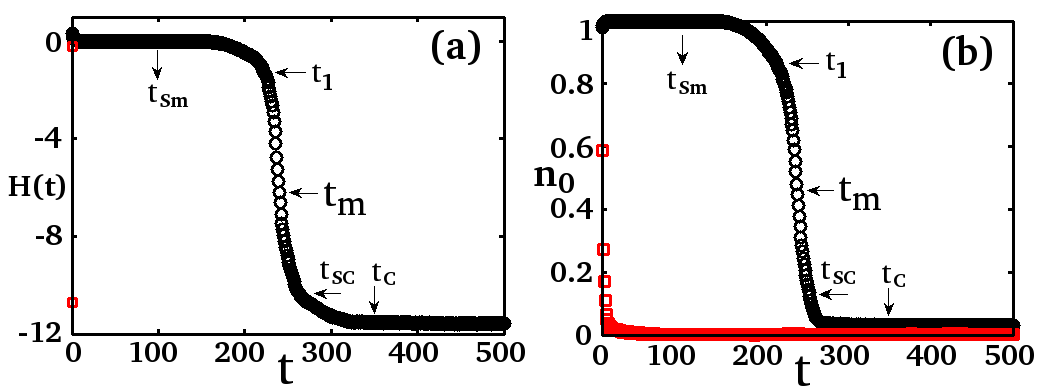}
\caption{{\it Energy and austenite fraction evolutions}: For $T=0.76, 0.4$ in quenched-temperature Regions 4 and 1 respectively,  the falloff of  a) the  Hamiltonian  energy, $ H(t)/ N$  vs $t$; and  b) the austenite fraction  $n_0 (t)\equiv 1 - n_m (t)$ vs $t$. }
\label{Fig.2}
\end{center}
\end{figure} 

\begin{figure}[h]
\begin{center}
\includegraphics[height=4.5cm, width=8.6cm]{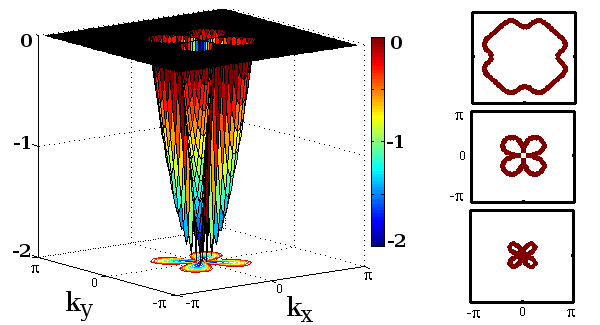}
\caption{(Color online) {\it Golf holes and funnels.}  Relief plot for $T= 0.76$ of  relevant martensitic spectrum $\epsilon (\vec k)$ versus $\vec k$ in the Brillouin zone, with a zero-energy plane. Sidebar: Golf hole edges $\epsilon (\vec k)=0$, for quench temperatures top to bottom,  that are low $T= 0.42$ (explosive  conversions);
medium $T= 0.76 < T_4$ (slow, rarer conversions);  and high $ T= 0.91 > T_4 = 0.824$ (no conversions, divergent entropy barriers).}
\label{Fig.3}
\end{center}
\end{figure}

 \begin{figure}[h]
\begin{center}
\includegraphics[height=2.9cm, width=8.7cm]{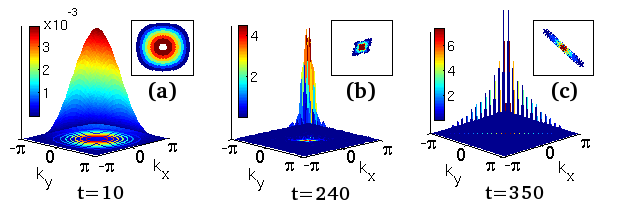}
\caption{(Color online) {\it Structure factors for domain-wall phases:}  Relief plots of $\ln [1 +|S (\vec k, t)|^2]$ versus $(k_x , k_y)$ in the Brilliouin Zone for a) vapour; b) liquid; c) crystal.
Insets: Corresponding contour plots.}
\label{Fig.4}
\end{center}
\end{figure}

Since the evolving energy  is  first zero, and then negative,  the relevant spectrum is  thus a  zero-energy plane  through the $\epsilon (\vec k)$ surface, plus  the negative  energies below it. Fig 3  shows that the resulting relief  plot naturally depicts a flat surface containing a  {\it golf hole} defined by $\epsilon(\vec k)  =0$, and a {\it funnel} $\epsilon (\vec k)  < 0$ inside it. The sidebar shows the temperature-dependent, anisotropic golf hole edge, that is large (small)  at low (high)  temperatures.There is  an outer (inner)   squared-radius of  $K^2 = {G_\pm} ^2\simeq 2/R_c (T)  \pm  \delta$ with $\delta \equiv  (A_1 {\bar V} /2 \xi_0 ^2) > 0$, where the average is $2/R_c \equiv - (g_L/\xi_0 ^2) - \delta ~ > ~0$.  

Protein folding is understood through concepts such as golf holes and funnels in configuration space \cite{R2,R3}. Here, we find such concepts appearing naturally in martensitic re-equilibration, but in a more easily represented form  in  the {\it label} space $\vec k$ of strain modes, in which the Hamiltonian is diagonal. For protein folding, this would correspond to the label space of folding normal-modes, of the protein model Hamiltonian.

 \begin{figure*}
\includegraphics[height=9.0cm, width=17.8cm]{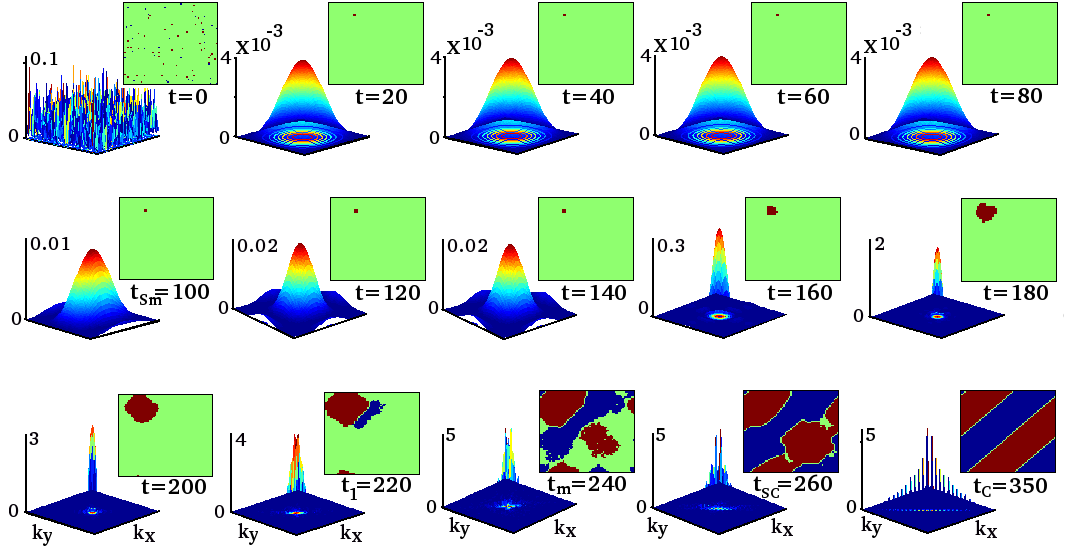}
\caption{(Color online) {\it  Textural evolutions: }   Snapshots of Fourier-space strain structure factors as  $\ln [1 +|S (\vec k, t)|^2]$ for successful conversions, after a  quench to $T= 0.76 $. Insets show coordinate-space  textures $\{S(\vec r, t)\}$, with green (blue/red) denoting austenite (martensite variants). See the movie. The martensite fraction conversion time is $t_m(T)$. Other crossover times $t_{Sm}, t_1, t_{SC}, t_C$ are defined in the text, and Fig 2. }
\label{Fig.5}
\end{figure*}

\begin{figure}[h]
\includegraphics[height=5.0cm, width=8.7cm]{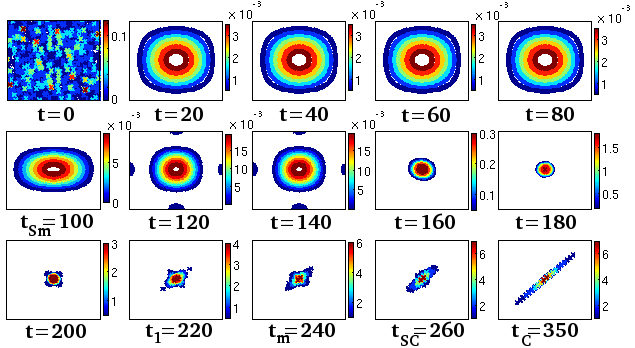}
\caption{{\it Contour plots  of structure factor in Fourier space:} Colour contour version of Fig 4 snapshots of   $ln [ 1 + |S({\vec k}, t)|^2 ]$ for $T= 0.76$. The three domain wall phases: vapour, liquid, crystal correspond to three contours: isotropic gaussian (e.g $t=80$), X-shaped (e.g $t=200$), single diagonal or fan-shaped (e.g $t= t_C \simeq 350$). } 
\label{Fig.6}
\end{figure}   

Fig 4 shows the Fourier-space structure factors $\{\rho (\vec k, t) \equiv |S(\vec k,t)|^2\}$ as relief plots,  for domain-wall phases of `vapour', `liquid', and `crystal'.
Fig 5  shows  the MC evolutions that transform the  phases into each other. 

We first discuss the insets of Fig 5, that are the evolving coordinate space textures $\{S(\vec r, t)\}$ as previously \cite{R12}, but now labelled by the characteristic times of Fig 2. See the movie in Supplementary Material. The insets show that the random seeds quickly form a vapour  droplet  of zero energy,  that fluctuates in place up to a time $t = t_{Sm} \sim 100$: it is in an `incubating' or ageing state \cite{R6,R8,R12}. The single-variant droplet then  finds and enters an energy-lowering, autocatalytic-twinning channel \cite{R17}  of  alternating opposite  variants, around a time $ t = t_1 \sim 220$. At a time $t=t_m \sim 240$,  a  domain-wall liquid forms, with  walls of wandering orientation. After a symmetry-breaking  choice of a diagonal at $t = t_{SC} \sim 260$, a domain wall crystal of oriented twins forms, beyond  $t = t_C \sim 350$. 

The main Fig 5 shows the same evolving textures, but now  in Fourier space. See also Fig 4 and Fig 6. For $t < t_{Sm}$, the ageing state $\rho (\vec k, t)$ for the incubating vapour droplet  persists unchanged as  a broad, isotropic  gaussian, poised over  the butterfly-shaped golf hole of Fig 3. One might expect the gaussian to promptly distort  along diagonals, to fit into the correspondingly anisotropic  golf hole, and narrow, to enter the negative energy funnel. Surprisingly, it does not do this.  For  $t > t_{Sm} \sim 100$,  it  waits in an incubation stage, to  develop wings along $k_x , k_y$ axes. See also the contour plots of Fig 6, at these times. Then for $t <  t_1$ the peak  narrows and then rises sharply, and for $t_m > t >  t_1$,   enters the golf hole, where it adopts the  bi-diagonal symmetry of  the funnel. After a symmetry-breaking at $t \sim t_{SC}$, the structure factor  at $t> t_C$ is  along a single diagonal. Thus the Fourier space distribution develops mis-oriented wings along the axes, before it forms wings along the Compatibility-favoured diagonals. 
In coordinate space, the austenitic $S=0$ spins at  the droplet surface must flip collectively to produce $S=\pm 1$ surface spin regions of the right symmetry: much like a kinetic constraint \cite{ R7}, but here  {\it self-generated}. The improbability of  finding this collective-spin distortion constitutes the entropy barrier.

Fig 6 shows the contour plots in the BZ corresponding to the relief plots of  the main Fig 5. The $\rho (\vec k, t)$ value of a point  on such contours represents the Fourier intensity or `occupancy', at a given  $\vec k$. It would also be interesting to monitor the  evolving occupancy at a given energy. 
 We  define  the energy occupancy distribution $\rho (\epsilon,t)$, similarly to that of a protein-folding simulation \cite{R4}, 
$$\rho(\epsilon, t) = \frac{\sum_{\vec k} \delta_{ \epsilon, \epsilon (\vec k)}  \rho(\vec k, t)}{ \sum_{\vec k} \rho(\vec k, t)}.~~(3.1)$$

\begin{figure}[h]
\begin{center}
\includegraphics[height=5.8cm, width=8.7cm]{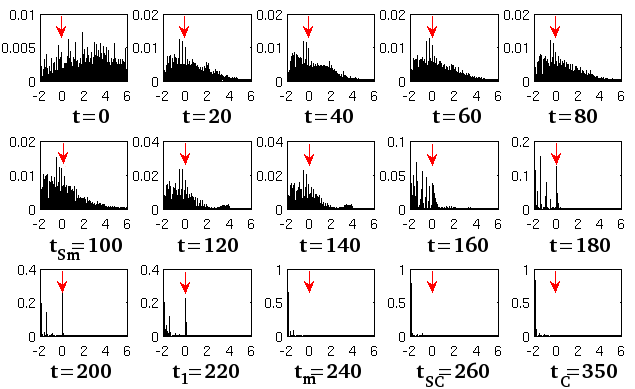}
\caption{{\it Time evolution of single-run energy occupancy distributions :} Plots of  $ \rho(\epsilon, t)$ versus $\epsilon$ for quench to $T= 0.76$  for a single run, at various times. Arrow marks the golf hole, of   energy $\epsilon =0$. The funnel region is $0 > \epsilon > g_L = -|g_L|$. A small peak at large energies appears  on the formation of the  entropically critical droplet for 
$t > t_{Sm} \simeq100$; and an occupancy spike from an entropy barrier persists at the golf hole energy right up to $t =t_{m} \simeq 240$. }
\label{Fig.7}
\end{center}
\end{figure}

\begin{figure}[h]
\begin{center}
\includegraphics[height=6.0cm, width=8.5cm]{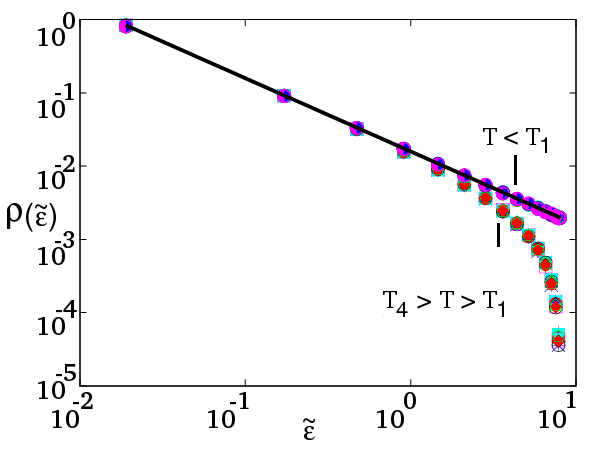}
\caption{{\it Final energy occupancy distributions:} Log-log plot of  $ \rho ({\tilde \epsilon}, t)$ versus $\tilde \epsilon \equiv \epsilon -g_L$ showing final  $1/\tilde \epsilon$ behaviour, regardless of quench temperature, and of different energy scales $E_0 = 3, 4, 5, 6$. Solid line has a slope  $-1.0$; and an estimated upper energy cutoff  \cite{R21} is $\simeq  9.1$, consistent with simulations. }
\label{Fig.8}
\end{center}
\end{figure}

Fig 7 shows the evolving histogram  of the {\it single-run}  $\rho( \epsilon, t)$ versus  energy $\epsilon$. The arrow marks the energy $\epsilon=0$ of the  golf hole edge. The  negative energies $0 > \epsilon > g_L$ are the funnel region.
The distribution remains fixed, up to $t=t_{Sm}=100$, and then  a small positive energy peak  appears, corresponding to the wings  along the axes of Fig 6.  (It  appears at different onset times in different runs, so a time average would wash it out.) The weight of the distribution moves more  into the negative energy region, and by around $t=t_m =240$, it is almost entirely in  the funnel. Note the long-lived, occupancy spike at the energy of the golf hole edge and its environs \cite{R3}.  This disappears at $t_m$, on crossing of the entropy barrier.

 As shown  in Fig 8  the final distribution, for all quenches, is an inverse-energy falloff in the excitation energy above the bulk Landau term, $\tilde \epsilon \equiv \epsilon - g_L > 0$:
 $$\rho( {\tilde \epsilon}, t; T) \rightarrow 1/{\tilde \epsilon}. ~~~(3.2)$$
   For a continuous-variable displacement of a harmonic oscillator, averaging with a Boltzmann factor would yield the same inverse-energy behaviour. The form is also independent  of Hamiltonian energy scales  $E_0$ and anisotropic stiffness constants $A_1$. Since the domain walls are sparse, and discrete, the energies are also discrete. The energy has an upper cut off, that is estimated \cite{R21} as  $\log_{10} {(\tilde \epsilon)} \lesssim  9.1$, consistent with the simulation results.

Notice that although the distribution $\rho (\vec k,t)$ or $\rho (\epsilon,t)$  of Fig 5 or  Fig 7 has much of its  weight poised above the funnel states inside the golf hole,
these domain-wall modes labelled by $\vec k$ or $\epsilon$, do not immediately collapse into available negative energy states. Re-equilibration does not follow a strategy of 'every mode for itself'. Rather  there is an `all modes together' strategy: the modes first partially equilibrate so there is  no net inter-mode energy exchange, setting up some                   nonequilibrium mode-distribution; followed by a slower emergence, as entropy barriers are crossed, of an equilibrium mode-distribution at the bath temperature.

\section{Conversion delays:  Entropy barrier crossing}
 
 We need to understand how conversion delays from entropy barriers, can be  so drastically different, at nearby temperatures. 
At low temperatures  of Region 1 of $T < T_1 =0.42$,  seeds in austenite convert explosively to martensite variants, for every run.
At  very high temperatures of Region 3 of $T _0 =1  > T > T_4 =0.82$ there is a complete `blocking'  of   conversion to energy-lowering martensite, with the zero-energy seeds dissolving for every run, into zero-energy austenite. And for  Region 4 of $ T_4 > T  > T_2 =  0.68 $   there is a rise in both the  conversion time  and the fraction of blocked runs on approaching $T_4$, so the mean time   diverges like a Vogel-Fulcher law\cite{R1,R12}  ${\bar t}_m (T) \sim e^{|\Delta S|}\sim e^{1/(T_4 - T)}$. 

An understanding of entropy barriers $|\Delta S|$ that are either zero, sharply rising, or infinite, comes from quenches to Region 4, with a  Fourier distribution $\rho (\vec k,t)$ that starts as an isotropic gaussian and ends as an inverted fan along one diagonal, as in Figs 4, 5. We parametrize the distribution through a weight $\eta$ of the emerging anisotropy.

The distribution  is separated, outside the golf hole ($t < t_1$),  and inside the funnel ($t > t_1$), as 
$$\rho (\vec k, t)  = \rho_g (\vec k, \eta) \theta (t_1 -t) + \rho_{f} (\vec k, \eta) \theta (t-t_1).~~(4.1)$$ 
 We will focus on the {\it constrained} search pathways outside the golf hole for $t < t_1$, through the parametrization

$$\rho_g=    N\{ n_{m}(0)(1 + b_g \eta (t)) + \eta (t)  \cos 4 \theta\}  C_g e^{- {{\vec K}^2}/ 2 \sigma^2}, ~~(4.2)$$
where $C_g$ normalises the gaussian to unity, and $b_g \equiv {n_{m}(0)}^{-1} -  {n_{m}(t_1)}^{-1}$.  The evolution parameter $\eta (t)$ then carries the fourfold anisotropy of the kernel of (2.1c), as   $\rho_g \sim \eta (t) ( 1 + \cos 4 \theta)$. Here the normalisation (2.4) yields $\eta (t) = [(n_m (t)/ n_{m}(0)) -1]/b_g  >0$.  The distribution is isotropic with $\eta (t) =0$, during the  $n_m (t) =n_{m}(0)$ incubation for $t_{Sm} > t$.  Whereas for  $t_1 >  t > t_{Sm}$ a nonzero $\eta > 0$  induces an  angular modulation, that  {\it increases}  the distribution at  $\theta = 0, \pi/2..$, i.e. along the $k_x, k_y$  axes. 

Writing the $\vec k \neq 0$ Hamiltonian energy of (2.1)  as averages $<...>$ over the distribution (4.2), so  
$$H/N n_m = \xi_0^2 <{\vec K}^2> + g_L + (A_1/2) <{\tilde V}(k)>, ~~(4.3a)$$
 we obtain on  the zero-energy plane outside the golf hole, a constraint linking the Ginzburg, Landau and St Venant contributions,
$$ H/Nn_m \xi_0^2 = \{ 2 \sigma^2 - 2/ R_c \}  +  \{\eta (t)/ n_m (t)\} \delta /2 =0. ~~~(4.3b) $$

In the $\eta =0$ ageing state the average golf hole radius determines the gaussian width or inverse droplet size as  $2\sigma^2=2/ R_c$. From  the constraint of (4.3b),  any  decrease in width must  be compensated by an {\it increase} in {\it mis}-orientation energy $\sim \delta \sim A_1 {\bar V}$ of the last term: the $\eta >  0$ wings must indeed, first emerge along the axes, before the  diagonals. This explains  the  observed $\rho (\vec k, t)$ distortions, of Figs 5,6  for $t_1 > t > t_{Sm}$. At $t =t_1$,  when $\eta (t_1) = n_{m}(t_1)$, the width from (4.3b) narrows to  $2\sigma^2 = (2/R_c) - \delta/2$; and then  to the inner radius. If $\eta$ is  (unphysically) taken to be negative, favouring diagonal wings right away, then the constraint of (4.3b) makes the width larger, going in the wrong direction.

The delays $t_m \sim e^{|\Delta S|}$ are understood through the temperature-dependent golf holes in the sidebar of Fig 3. For $T$  below $T_1$, the golf hole in the BZ is large, and the flat distribution from seeds directly  forms a liquid distribution of  Fig 4, entering the funnel immediately for every run. The conversion time is negligible, and its  entropy barrier is zero, $|\Delta S| \simeq 0$.  For $T$  approaching $T_4$ the golf hole shrinks,  and hence the search times rise;  the entropy barrier diverges as $|\Delta S| \sim 1/ (T_4 - T)$, and the fraction of runs converting to martensite falls, yielding Vogel-Fulcher behaviour. For $T > T_4$ the golf hole inner radius  $G_- ^2$ closes, and the  resulting 4-petalled golf hole topology presents the isotropic gaussian with an infinite entropy barrier. Thus even though the martensite Landau energy  is lower  than  the austenite energy for $T_0 > T > T_4$, it becomes ergodically inaccessible \cite{R8} to the  small and dilute initial seeds. 

For nucleation by activation over {\it energy} barriers, a divergent droplet  timescale is associated with a {\it divergent size} in coordinate space. By contrast, for non-activated {\it entropy} barrier crossing,  a divergent search timescale for droplet pathways  is associated with a {\it shrinking  bottleneck} in Fourier space.

 Under MC dynamics for a given run $n$,  the total  Hamiltonian energy $H (t)$  of a texture $\{S(\vec r,t)\} $ goes to $H(t+1)$  at the next MC sweep. The probability for a given energy change $\Delta E$ to occur at a time $t$, from Hamiltonian increments $\Delta H (t,n)= H(t+1) - H(t)$, is obtained through an average over all runs $n =1,2...N_{runs}$: 
$$P(\Delta E, t) \equiv \frac{1}{N_{runs}} \sum_{n=1}^{N_{runs}} \delta_{\Delta H(t,n), \Delta E}.~~(4.4a)$$
At early times, the probability is peaked at negative values $\Delta E = - |\Delta E| <0$, with an asymmetric shoulder on the negative side \cite{R7}; and  at long times, this becomes an equilibrium  distribution, symmetric around zero (not shown). 
To determine  if there is a dominant  energy change during the evolution, regardless of when it occurs, we average the  
 energy release probability  over the entire holding time, $P(\Delta E) \equiv \sum_t P(\Delta E, t) / t_h$. 

 Fig 9 shows $P(\Delta E)$  for various temperatures  and energy reductions. There are a few large magnitudes of energy release, but mostly, $P(\Delta E)$  falls as a powerlaw in the magnitudes $\sim 1/ |\Delta E|^\gamma$, with  a common exponent $\gamma \simeq 2$. This suggests that the  domain-wall adjustments have no characteristic energy scale, and are like small earthquakes, of all scales. Acoustic emissions occur in martensites, from twin boundaries inducing energy changes, and power law distributions have been seen \cite{R22} with  exponent close to $2.3$. 

\begin{figure}[h]
\begin{center}
\includegraphics[height=5.8cm, width=8.5cm]{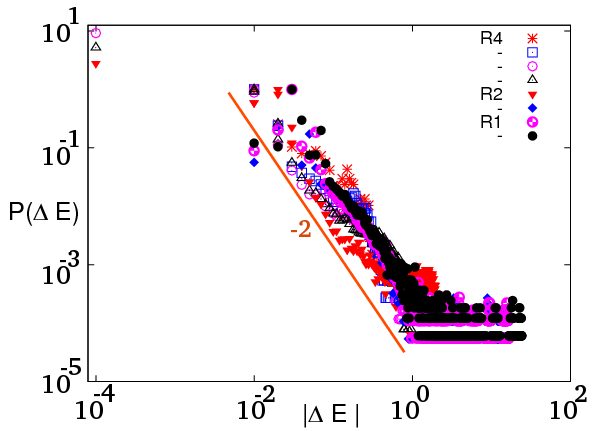}
\caption{ {\it  Probability distribution  of energy releases:} Log-log probability of energy releases at any time, $P(\Delta E)$,  versus  magnitude of releases $|\Delta E|$. }
\label{Fig.9}
\end{center}
\end{figure}  
The Monte Carlo acceptance fraction\cite{R6}  $A_{acc} (t)$,  is the fraction of $N$ sites where the MC move is {\it accepted}  during a given sweep at $t$. Ageing non-interacting oscillators, have  small and monotonically decreasing \cite{R6}  $A_{acc} \sim 1/t \ln t$, from an inefficient, memory-less  search of all oscillators, for  an ever-decreasing un-relaxed population. 
 
Figure 10 shows  for this model, the very different  acceptance fractions  $A_{acc}(t)$ versus time $t$ for  the three Regions.
 At $T = 0.76$ in Region 4, $A_{acc} (t)$ is nearly zero during incubation. The acceptance spikes at conversion times $t= t_m$ during conversion from domain-wall  vapour to liquid,  and falls again to zero  beyond $t= t_C$, in the crystal phase. The spike occurs at the same time as the  sharp rise of the martensite fraction  through $n_m (t_m) =0.5$, giving a physical justification to this earlier definition of $t_m$.
 
 \begin{figure}[h]
\begin{center}
\includegraphics[height=2.5cm, width=8.7cm]{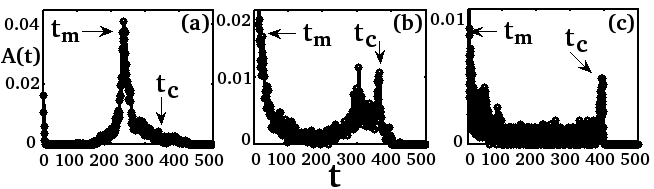}
\caption{{\it  Monte Carlo Acceptances :} Plot of $A_{acc}(t)$ versus  $t$ for  the three temperature-quench  Regions 4,
2, 1.  a)  $T=0.76$ in Region 4; b)  $T= 0.55$ in Region 2; c) $T =0.4$ in Region 1.  }
\label{Fig.10}
\end{center}
\end{figure} 
For $T =0.55$ in Region 2, $A_{acc} (t)$ is high initially, accepting most of the flips; decreases when the domain wall liquid phase is reached;
peaks again near $t= t_C$ on domain wall orientations; and finally  falls to zero acceptances  in the crystal phase. There can be a second peak just  before $t=t_C$, where austenite droplets are generated to bind to the domain walls \cite{R12}. For $T= 0.4$ in Region 1, the acceptance  in the liquid phase is  spiky, during domain wall motion in the liquid phase. Then there is a peak in $A_{acc} (t)$, as a large number of austenite droplets  or hotspots are generated to catalyse domain-wall symmetry-breaking orientations \cite{R12}, finally  falling to zero as before,  in the  crystal.  
 
 The ageing state has MC acceptance fractions that are nearly zero, and acceptance  spiking at $t_m$ or $t_C$ is  thus a diagnostic at all temperatures for the crossing of an entropy barrier.

 \section{ Domain-wall thermodynamics and effective temperatures}

We approximate the MC Hamiltonian $H(t)$ by that of independent spins at $T$, in a time-dependent local  mean-field \cite{R12}   $\sigma (\vec r, t)$, where the non uniformity comes from the domain walls. This implicitly assumes (consistent with simulations) that there is a separation of time scales, with individual spins flipping rapidly in response to quenches of the temperature, with domain wall configurations evolving more slowly. The Appendix obtains, within a `time-dependent local mean-field' approximation, expressions for the free energy $F \simeq F_{LMF} (t)$, internal energy $U (t)$, and entropy  $S_{entr} (t) =  -F_{LMF} (t) + T S(t)$, in terms of the $\{ S(\vec r,t)\}$ configurations at a given MC sweep  labelled by $t$.  

We regard the domain-walls under stress as {\it internal work sources}, that run freely after a quench. The overall change in the internal energy  $d U (t)$, by a First Law type relation, is a sum of contributions from  the work done by domain walls ${\dbar}W(t)$, and  the heat release $\dbar Q (t)$: 
$$ dU (t) = \dbar W(t) + \dbar Q (t).~~ (6.1)$$

We need relations between the textural thermodynamics and  increments in the  heat and  the work,  at constant $T$. The heat release  by spins, that are at  the bath temperature, is  taken as $\dbar Q (t)= TdS_{entr} (t)$. For equilibrium, the Helmholtz free energy change between thermodynamic states, is the available work  at constant temperature \cite{R23}. 
We assume the free-running  work  increment at constant temperature saturates a similar availability, set by the  evolving free energy  change: $\dbar W(t) = d F_{LMF} (t)$. At long times after entropy barriers are crossed,  thermodynamic equilibration $dF =0$ is accompanied by mechanical equilibration  $\dbar W=0$.  

The evolving work rate is ${\dot W} (t)$, and  the heat emission rate is ${\dot Q}(t)$, where `rates' are $X(t +1) - X (t) \equiv {\dot X } (t)$.  Fig 11 shows the probability distributions $P(\dot W), P(\dot Q)$ over MC runs,  for the rates of internal work done or heat emitted. They can peak at negative values, but finally both equilibrate to peaks centred at zero. Fig 12 shows  that the mean rates of work done and heat emitted are suppressed in the ageing state  by entropy barriers, but show  large and sudden releases, as the entropy barriers are crossed. On equilibration, all mean rates tend to zero.

\begin{figure}[h]
\begin{center}
\includegraphics[height=5.0cm, width=8.5cm]{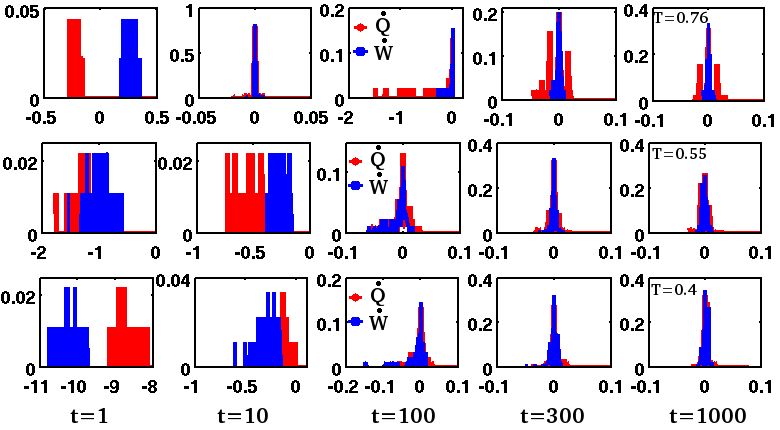}
\caption{{\it  Evolution of distributions for rate of heat  emission and internal work  :} Plot of  evolving $ P(\dot Q, t)$ versus  $\dot Q$  and  $P(\dot W, t)$ versus  $\dot W$ for quenches into the three  Regions 4, 2, 1. {\it Top row:} $T=0.76$ in Region 4. {\it Middle row:} $T= 0.55$ in Region 2. {\it Bottom row:} $T =0.4$ in Region 1. For all $T$, the final distributions are peaked symmetrically around zero. }
\label{Fig.11}
\end{center}
\end{figure}

\begin{figure}[h]
\begin{center}
\includegraphics[height=2.8cm, width=8.7cm]{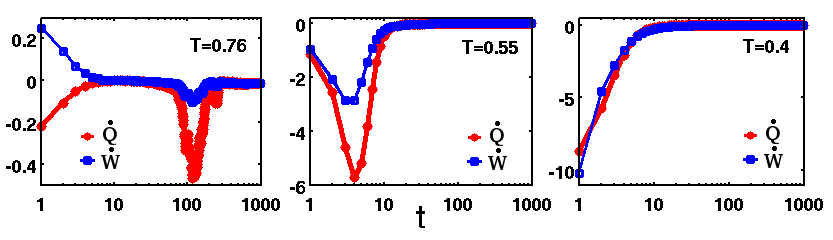}
\caption{{\it Evolution of averaged rates of heat emission and internal work: }  Log-linear plots shows the mean rates $<\dot W>$ and $<\dot Q>$ versus the time $t$. For $T= 0.76$,  the rates are zero in the ageing state, have large releases near ${\bar t}_m (T)$,  and finally vanish on  equilibration. For  $T = 0.55, 0.4$  the curves are similar, but moved to the left, as conversion delays $t_m$, vanish. For all $T$, the final mean rates are zero. }
\label{Fig.12}
\end{center}
\end{figure}

The  energy of trapped stresses can only escape as heat  released to the bath, since  the boundary conditions are periodic, and not piston-like. One would like to relate a  stress-induced heat release to re-equilibration of some {\it effective temperature}, that  has otherwise been defined in terms of the Fluctuation-Dissipation relation \cite{R24}.

In  the equilibrium case, and for some  externally imposed work protocol \cite{R25},  one can distinguish\cite{R26}  between heat changes (that are occupancy changes  of given energy levels), and work changes (that are energy  levels  changes at fixed-occupancy). For temperature quenches, however, there is no such external sequential  control, and both work and heat changes occur together.  The  relative proportion of spontaneous heat and work in (6.1)  can be tracked, by defining a $T_{eff} (t)$,through
$$\dbar W (t) \equiv [1 - \frac{T}{ T_{eff} }] d U (t); ~~~ \dbar Q (t) \equiv \frac{T}{ T_{eff} } dU (t).~~~(6.2)$$
With the previous increment relations this is manifestly equivalent to  a 'microcanonical' definition,  $T/T_{eff}(t)  = Td S_{entr} (t) /d U (t)$, as invoked in protein folding models \cite{R4}.

The incremental work can also be related to changes in the local mean-field,
$$\dbar W(t) \simeq - \sum_r p (\vec r, t)\delta \sigma(r,t)~~(6.3)$$
where the local stress is $p(\vec r,t)= \partial F _{LMF} (t)/\partial \sigma (r,t)$. For  static textures at equilibrium $\sigma(r,t) ={\bar \sigma }(\vec r)$ satisfying the mean-field self-consistency condition \cite{R27}, the local stresses vanish  in the domain-wall crystal phase \cite{R12}, as seen in the Appendix.
 From $\dbar W \sim (1- T/ T_{eff})$ of (6.2),  this vanishing of trapped stresses is  consistent with  $T_{eff} (t) \rightarrow T$.  A detailed study of this final equilibration would involve the second entropy barrier at $t_C$, and  so the trapped-stress related effective temperature will be pursued elsewhere.

 \section{Summary}

We develop a detailed understanding of the re-equilibration process  of  domain walls, in a  martensite-related three-state model with powerlaw anisotropic interactions. There is a natural appearance of concepts borrowed from  protein folding, of golf holes and funnels; and from oscillator relaxation models, of Monte Carlo acceptance fractions. As found earlier, domain-wall  phases after a quench, evolve from a domain-wall 'vapour'   to a 'liquid', and thence to a `crystal' of oriented walls. There is a temperature regime where the martensite conversion delay $t_m$ dominates the total delay. The  evolution  from  a vapour-phase zero-energy droplet in zero-energy austenite to a  negative-energy  liquid of martensite domain walls, is best understood in Fourier space. The  droplet  has an isotropic  gaussian structure factor, peaked at the Brillouin zone centre, over  a butterfly-shaped, small golf hole with a negative energy funnel inside. The incubation delays come from a search for  anisotropies  at  zero energy to roll  into the golf hole. At low temperatures, the golf hole is large, and conversion from seeds occurs  almost immediately. At temperatures above transition, there is  golf hole topology change, preventing the roll-in, and suppressing the conversion of austenite.

In the ageing state, the MC acceptance fractions are negligible;  both work and heat rates are zero; and trapped local stresses  are held in place  by entropy barriers. On crossing entropy barriers at the death of ageing, there are  spikes in the MC acceptance, and sudden releases of the trapped stress and  heat. On the resultant thermal and mechanical re-equilibration,  acceptance fractions for spin-flips again vanish;  work and heat rates are zero; and  the oriented domain walls are stress-free, with a related effective temperature going to the bath temperature. The scenario may have relevance to other models of athermal re-equilibration after a  quench, such as glassy and granular models.\\

It is a pleasure to thank Shamik Gupta, Uwe Klemradt, Stefano Ruffo, VSS Sastry and Sandro Scandolo for useful conversations. NS thanks the University Grants Commission, India  for a Dr. D.S. Kothari postdoctoral fellowship.\\ \\ \\

{\bf{ Appendix :  Time-dependent local mean-field approximation :}}\\

The {\it uniform}, static   mean-field approximation is familiar for ferromagnets, and antiferromagnets (where it is a staggered magnetisation). However a static,  {\it local} mean-field can faithfully reproduce the domain-wall textures from simulations\cite{R27}. This can be generalized \cite{R12}  to an {\it  MC time-dependent} local mean-field approach, to describe the evolving domain walls.

The model Hamiltonian  of (2.1) has the bath temperature $T$ entering  $H(T)$ through the martensitic strain magnitudes ${\bar \varepsilon} (T)$. 
With the partition function as $Z=\sum_{\{ S\}} e^{ -H(T)/T}$, 
and the canonical free energy as $ F = -T\ln Z$, the entropy is $S_{entr} = -\partial F/\partial T$, 
the internal energy  defined by $U \equiv F + T S_{entr}$, is then not just the averaged hamiltonian, but is 
$$U = < H> - <T \partial H/\partial T>.~~~(A1)$$
 
With strain-pseudospin patterns  $\{ S(\vec r,t)\}$ evolving under Monte Carlo dynamics, the weight factor $e^{-H /T}$is truncated within \cite{R12} a
`time-dependent  local mean-field approximation'  that we restate here for completeness. It is defined by the substitution into  the coordinate space Hamiltonian as 
$$S(\vec r) S (\vec r') \rightarrow S (\vec r) \sigma({\vec r } ', t)  +   \sigma (\vec r, t)S (\vec r)  - \sigma (\vec r,t) \sigma ({\vec r}',t). ~(A2a) $$
Here, the mean field spin $\sigma$ is defined as a run-average of a local spin variable at a site $\vec r$, and an MC  time $t$: 
$$\sigma (\vec r,t) \equiv < S(\vec r,t)> .~~(A2b)$$ 
The local mean-field  weight  is then $e^{-H/T} \rightarrow  e^{-H_{LMF}/T}$ where
$$H_{LMF} /T = \sum_{\vec r} q(\vec r,t) S(\vec r,t) -\frac{1}{2} \sum q (\vec r,t) \sigma (\vec r, t)~~~(A2c)$$
depends on individual spins in a mean-field, and 
$$q(\vec r,t) \equiv  \sum_{\vec r'} q_0 (\vec r -\vec r') \sigma (\vec r',t),~~(A2d)$$
where
$$q_{0} (\vec r -\vec r') \equiv D_0 [\{ g_L(\tau)  +  \xi_0^2 { \Delta_{\vec r}}^2\} \delta_{\vec r, \vec r'}  + \frac{A_1}{2} V (\vec r - \vec r' )].~(A2d)$$
As mentioned in the text, the  individual spins are assumed to respond instantaneously to 
the quenched temperature of the heat bath and to the influence of domain walls, that themselves  evolve much more slowly, under MC dynamics. 

The  corresponding substitution in the  Fourier space Hamiltonian  of (2.1) is
$$|S(\vec k)|^2 \rightarrow S(\vec k) \sigma (\vec k, t)^* + \sigma (\vec k, t) S(\vec k, t)^* - |\sigma (\vec k, t)|^2.~~(A3a)$$
Here
$$H_{LMF} /T = \sum_{\vec k} q(\vec k,t)^* S(\vec k,t) -\frac{1}{2} q_0 (\vec k, t) |\sigma (\vec k, t)|^2,~~~(A3b)$$
 where
$$q(\vec k,t) \equiv q_0 (\vec k) \sigma (\vec k,t),~~(A3c)$$
 and 
$$q_{0} (\vec k) \equiv D_0 [\{ g_L(\tau) +  \xi_0^2 { \vec K}^2  \}  + \frac{A_1}{2} {\tilde V} ({\vec k} )(1-\delta_{\vec k, 0})]~~(A3d)$$
with $D_0 (T) \equiv  K_0/T =2 \bar{\varepsilon}^2 (\tau)  E_0/ T$.

The thermodynamic functions  are  all taken as zero in $S=0$ uniform austenite,  and the approximate $LMF$ free energy is
$$ F_{LMF} =  - T\{ \sum_{\vec r}  \ln \frac{1}{3} [ 1 + 2 \cosh q(\vec r)] - \frac{1}{2} \sum_{\vec k}  q_0 |\sigma(\vec k)|^2\}, ~(A4)$$
The internal stress $p (\vec r,t) = \partial F_{LMF} / \partial \sigma (\vec r,t)$ is
$$p(\vec r,t) =- \sum_{{\vec r}'}q_0 (\vec r - \vec r')[ \frac{2 \sinh q (\vec r ',t)}{1+ 2 \cosh q (\vec r ', t)} + \sigma (\vec r ', t)]~~(A5)$$
and vanishes at the self-consistent, static equilibrium textures \cite{R27}  $\sigma (\vec r, t) ={\bar \sigma} (\vec r)$, when the square bracket is zero.

From (A1) and the Hamiltonian (2.1), the internal energy is

$$U = [ 1 - \frac{T }{{\bar \varepsilon}^2}  \frac{d {\bar \varepsilon}^2}{dT}] < H> -E_0 { \bar \varepsilon}^2T \frac{d g_L}{d T} \sum_{\vec k} < |\sigma(\vec k)|^2>,~~(A6)$$
where the averages are with the $LMF$ weight.  The entropy is taken as the difference of (A4), (A6) 
$$TS_{entr} = - F_{LMF} +U.~~(A7)$$
With $Td/ dT = [\tau (T) - \tau(0)] d/ d\tau$  and the definitions of the text,
$Td{\bar \varepsilon^2}/ dT = -[ \tau (T) -\tau (0)]/ (4 \sqrt{1- 3\tau/4})$; and 
$Td g_L/dT = [\tau(T) -\tau (0)]  [ 1 -2{(\bar \varepsilon}^2 -1)/(4 \sqrt{1-3\tau/4})].$

Of course, these expressions are in terms of the time-dependent local mean-field $\sigma (\vec r,t)$. We sidestep the evaluation through (A2b) of $\sigma (\vec r,t)$ at each time $t$, by invoking  the spirit of mean-field approximations, namely  that \\
`the function of an average is an average of the function', and taking
$$F_{LMF} (\{\sigma (\vec r,t)\} )\simeq < F_{LMF} (\{S (\vec r,t)\})>;~~~(A8a)$$
$$U (\{\sigma (\vec r,t)\} )\simeq < U (\{S (\vec r,t)\})>;~~~(A8b)$$
where the averages are now taken over each distinct MC runs.

Thus the (time-dependent)  $LMF$ expressions of (A4), (A6), (A7) yield expressions for the domain-wall thermodynamics of an evolving texture $\{S(\vec r,t)\}$ over a each distinct re-equilibration run,  
that can then be averaged over many runs. This approach has been used for Figs 11,12.


\begin{thebibliography}{2004}
\bibitem{R16}
 Instead of  an isothermal martensite  regime with activated conversion times\cite{R12} $\ln t_m \sim E_0/T$, we work  in an  athermal regime  where non-activated conversion times are insensitive to energy scales $E_0/ T_0 > 1$. At $t=0$, the  $S = 0$ austenite background, seeded with $S = \pm 1$ variants, at $2 \%$  of randomly located sites, is   quenched to  $T $, that is  held fixed   for  $t  \leq t_h$  MC sweeps.  
Typical scaled parameters are \cite{R12} $T_c  = 0.9,  \xi_0 = 1; ~A_1 = 4; ~2 A_1 / A_3 = 1; ~E_0 =3,4,5,6$. Here$~N=  64^2$; the holding time is  $t_h \leq 10^4$; and random-seed averages are over $N_{run} =100$ runs. For successful runs converting at a time $t_m < t_h$, the rate is $r\equiv 1/t_m$,  and the inverse average rate  defines  a mean conversion time ${\bar t}_m \equiv 1/<r>$. (For runs where seeds vanish into austenite, and never return within $t_h$,   we set $r =1/t_h$.) Close to $T= T_4$, when conversions become rarer, we find \cite{R12}  a Vogel-Fulcher behaviour  ${\bar t}_m \sim e^{(T_4 - T_1)/(T_4 - T)}$, and a log-normal distribution for $T_{4} > T > T_{2}$,  insensitive to $t_h$ and to $E_0$.
\bibitem{R18}
The Fourier amplitude $S(\vec k,t)$ is not just an independent  oscillator displacement  labelled by $\vec k$, since its (inverse) Fourier transform is constrained to be an integer $S(\vec r, t) =0,\pm 1$.
\bibitem{R21}
The cutoff energy is  the maximum value of  $\tilde \epsilon (\vec k)$. Estimating the two terms for  diagonal or axis directions of  $\hat k$, the upper bound on is  ${\tilde \epsilon}  < Max (8 \xi_0^2, 4 \xi_0 ^2+\frac{A_1}{2})$. For $\xi_0^2=1$, $A_1=4$, estimated cutoff is $\log_{10} (\tilde \epsilon) \lesssim 9.1$, consistent with simulations.
\end{thebibliography}
\end{document}